\documentclass[preprint,aps,showpacs,preprintnumbers,amsmath,amssymb,nofootinbib]
{revtex4}
\usepackage{epsfig}
\begin{document}

\begin{flushright}
%\preprint{hep-ph/xxyyzz}
%\today\\
%UH-0528
\end{flushright}

%%%%%%%%%%%%%%%%%%%%%%%%%%%%%%%%%%%%%%%%%%%%%%%%%%%%%%%%%%%%%%%%%%%%%%%%

\newcommand{\be}{\begin{equation}}
\newcommand{\ee}{\end{equation}}
\newcommand{\bea}{\begin{eqnarray}}
\newcommand{\eea}{\end{eqnarray}}
\newcommand{\nn}{\nonumber}

\def\lb{\Lambda_b}
\def\ll{\Lambda}
\def\mb{m_{\Lambda_b}}
\def\ml{m_\Lambda}
\def\s1{\hat s}
\def\ds{\displaystyle}

%%%%%%%%%%%%%%%%%%%%%%%%%%%%%%%%%%%%%%%%%%%%%%%%%%%%%%%%%%%%%%%%%%%%%%%

\title{\large Fourth generation effect on $\Lambda_b$ decays}
%\author{R. Mohanta }
%\affiliation{ School of Physics, University of Hyderabad, Hyderabad
%- 500 046, India}
\author{R. Mohanta$^1$, A. K. Giri$^2$ }
\affiliation{
$^1$ School of Physics, University of Hyderabad, Hyderabad - 500 046, India\\
$^2$ Department of Physics, Indian Institute of Technology
Hyderabad, Yedumailaram - 502205, Andhra Pradesh, India}

\begin{abstract}
The rare decays of $\Lambda_b$ baryon governed by the quark level
transitions $ b \to s$,  are investigated in the fourth quark
generation model popularly known as SM4. Recently it has been shown
that SM4, which is a very simple extension of the standard model,
can successfully explain several anomalies observed in the CP
violation parameters of $B$ and $B_s$ mesons. We find that in this
model due to the additional contributions coming from the heavy $t'$
quark in the loop, the branching ratios and other observables in
rare $\Lambda_b$ decays deviate significantly from their SM values.
Some of these modes are within the reach of LHCb experiment and
search for such channels are strongly argued.

\end{abstract}

\pacs{13.30.Eg, 13.30.Ce, 12.60.-i} \maketitle

\section{Introduction}
The rare decays of $B$ mesons involving flavor changing neutral
current (FCNC) transitions are of great interest to look for
possible hints of new physics beyond the standard model (SM). In the
SM, the FCNC transitions arise only at one-loop level, thus
providing an excellent testing ground to look for new physics.
Therefore, it is very important to study FCNC processes, both
theoretically and experimentally, as these decays can provide a
sensitive test for the investigation of the gauge structure of the
SM at the loop level. Huge experimental data on both exclusive and
inclusive $B$ meson decays \cite{hfag} involving $ b \to s$
transitions have been accumulated at the $e^+ e^-$ asymmetric $B$
factories operating at $\Upsilon(4S)$, which motivated extensive
theoretical studies on these mesonic decay modes.

Unlike the mesonic decays, the experimental results on FCNC mediated
$\Lambda_b$ baryon decays e.g., $\Lambda_b \to \Lambda \pi $,
$\Lambda_b \to p K^-$, $ \Lambda_b \to \Lambda \gamma$ and
$\Lambda_b \to \Lambda l^+ l^-$ are rather limited. At present we
have only upper limits on some of these decay modes \cite{pdg}.
Heavy baryons containing a heavy $b$ quark will be copiously
produced at the LHC. Their weak decays may provide important clues
on flavor changing currents beyond the SM in a complementary fashion
to the $B$ decays. A particular advantage of the bottom baryon
decays over the $B$ mesons is  that these decays are self-tagging
processes  which should make their experimental reconstructions
easier.

Another important aspect is that, in the past few years we have seen
some kind of deviations from the SM results in the CP violating
observables of $B$ and $B_s$ meson decays involving $b \to s$
transitions \cite{hfag,utfit,browder,soni10, lenz10}. Several new
physics scenarios are proposed in  literature to account for these
deviations \cite{np}. Therefore, it is quite natural to expect that
if there is some new physics present in the $b \to s$ transitions of
$B$ meson decays it must also affect the corresponding $\Lambda_b$
transitions. Therefore, the study of the rare $\Lambda_b$ decays is
of utmost importance to obtain an unambiguous signal of new physics.

In this paper we would like to study the rare $\Lambda_b$ decays in
a model with an extra generation of quarks, usually known as SM4
\cite{4gen}. SM4 is a simple extension of the standard model with
three generations (SM3) with the additional up-type ($t'$) and
down-type ($d'$) quarks. The model retains all the properties of
SM3. The $t'$ quark like the other up-type quarks contribute to the
$b \to s$ transition at the loop level. Due to the additional fourth
generation there will be mixing between the $b'$ quark  the three
down-type quarks of the standard model and the resulting mixing
matrix will become a $4 \times 4$ matrix ($V_{CKM4})$. The
parametrization of this unitary matrix requires six mixing angles
and  three phases. The existence of the two extra phases provides
the possibilities of extra source of CP violation. Another advantage
of this model is that the heavier quarks and leptons in this family
can play a crucial role in dynamical electroweak symmetry breaking
as an economical way to address the hierarchy problem \cite{ewsb}.
The effect of fourth generation of quarks in various $B$ decays are
extensively studied in the literature \cite{4thgen}. In Refs.
\cite{rm1, buras}, it has been shown that this model can easily
explain the observed anomalies in the $B$ meson sector.

The paper is organized as follows. In section II we discuss the
nonleptonic decay of $\Lambda_b $ baryon. The radiative decay
process $\Lambda_b \to \Lambda \gamma$ is discussed in section III.
The results on semileptonic decays are presented in section IV.
Section V contains the summary and conclusion

\section{Decay width of $\Lambda_b \to \Lambda \pi^0$ and $\Lambda_b \to p K^-$ modes}

In this section we will discuss the nonleptonic rare $\Lambda_b$
decay mode $\Lambda_b\to \Lambda   \pi $ and $\Lambda_b \to p K^-$
induced by the quark level transition $ b \to s q \bar q, ~
(q=u,d)$. The effective Hamiltonian describing these processes is
given by \cite{hycheng} \bea {\cal H}_{eff}=\frac{G_F}{\sqrt 2}\left
[ V_{ub} V_{us}^* \sum_{i=1,2} C_i(\mu) O_i - V_{tb} V_{ts}^*
\sum_{i=3}^{10} C_i(\mu) O_i \right ], \eea where $C_i(\mu)$'s are
the Wilson coefficients evaluated at the renormalization scale
$\mu$, $O_{1,2}$ are the tree level current-current operators,
$O_{3-6}$ are the QCD and $O_{7-10}$ are electroweak penguin
operators.

Let us first consider the decay process $\Lambda_b \to \Lambda \pi$.
In the SM this mode receives contributions from the color-suppressed
tree and the electroweak penguin diagrams and the amplitude for this
process in the factorization approximation is given as \cite{rm2001}
\bea {\cal A}(\Lambda_b(p) \to \Lambda (p') \pi^0(q)) & =
&\frac{G_F}{\sqrt 2} \Big[
V_{ub}V_{us}^* a_2 - V_{tb} V_{ts}^* \left ( \frac{3}{2}(a_9 -a_7) \right ) \Big]\nn\\
& \times & \langle \Lambda (p') |(\bar s \gamma^\mu (1-\gamma_5) b |
\Lambda_b(p) \rangle \langle \pi^0 (q) | \bar u \gamma_\mu (1-
\gamma_5) u | 0 \rangle , \eea where $a_i= C_i+ C_{i+1}/N ~(C_i+
C_{i-1}/N)$ for $i$= odd (even). In order to evaluate the matrix
elements we use the following form factors and decay constants. The
matrix elements of the various hadronic currents between initial
$\lb$ and the final $\ll$ baryon, are parameterized in terms of
various form factors \cite{cdlu} as \bea \langle \ll (p') |\bar s
\gamma_\mu b | \lb(p) \rangle & =& \bar u_\ll (p') \Big[ g_1(q^2)
\gamma_\mu + i g_2 (q^2)\sigma_{\mu \nu} q^\nu +g_3 (q^2)q_\mu \Big]
u_{\lb}(p)\;,\nn\\
\langle \ll (p')|\bar s \gamma_\mu \gamma_5 b | \lb (p)\rangle & =&
\bar u_\ll(p') \Big[ G_1(q^2) \gamma_\mu + i G_2 (q^2)\sigma_{\mu
\nu} q^\nu + G_3 (q^2) q_\mu \Big]\gamma_5
u_{\lb}(p)\;,\label{matrix} \eea where $g_i ~(G_i)$'s  are the
vector (axial vector) form factors and $q$ is the momentum transfer
i.e., $q=p-p'$. The matrix element $\langle \pi(q) | \bar u
\gamma_\mu \gamma_5 u | 0 \rangle $ is related to the pion decay
constant $f_\pi$ as \be \langle \pi^0(q) | \bar u \gamma^\mu
\gamma_5 u | 0 \rangle = i f_\pi q^\mu/\sqrt{2}. \ee With these
values one can write the transition amplitude for $\Lambda_b \to
\Lambda \pi $ as \bea {\cal A}(\Lambda_b \to \Lambda \pi^0) &= & i
\frac {G_F}{2} f_\pi \left ( V_{ub} V_{us}^* a_2 -
\frac{3}{2}V_{tb}V_{ts}^*(a_9 - a_7)
 \right )\nn\\
& \times & \bar u_{\Lambda}(p')\biggr [ \Big(g_1(q^2) (m_{\Lambda_b}-m_\Lambda ) + g_3(q^2) m_\pi^2 \Big)\nn\\
&+&
 \Big(G_1(q^2)(m_{\Lambda_b} + m_{\Lambda}) -G_3(q^2) m_\pi^2 \Big)\gamma_5 \biggr] u_{\Lambda_b}(p).
\eea The above amplitude  can be symbolically written as \bea {\cal
A}(\Lambda_b(p') \to \Lambda(p) \pi^0(q)) = i \bar u_{\Lambda} (p')
(A + B \gamma_5) u_{\Lambda_b}(p)\;, \eea where $A$ and $B$ are
given as \bea A &=& \frac{G_F}{2}f_\pi \left ( V_{ub} V_{us}^* a_2 -
\frac{3}{2}V_{tb}V_{ts}^*(a_9 - a_7)
 \right )
 \times   \Big(g_1(q^2) (m_{\Lambda_b}-m_\Lambda ) + g_3(q^2) m_\pi^2 \Big),\nn\\
B &= &\frac{G_F}{2}f_\pi \left ( V_{ub} V_{us}^* a_2 -  \frac{3}{2}V_{tb}V_{ts}^*(a_9 - a_7)
 \right )\times
 \Big(G_1(q^2)(m_{\Lambda_b} + m_{\Lambda}) -G_3(q^2) m_\pi^2 \Big).
\eea Thus, one can obtain the decay width for this process as
\cite{pakvasa}, \bea \Gamma = \frac{p_{cm}}{8 \pi} \left [
\frac{(m_{\Lambda_b}+m_\Lambda)^2-m_{\pi}^2}{m_{\Lambda_b}^2}|A|^2
+\frac{(m_{\Lambda_b}-m_{\Lambda})^2 -
m_\pi^2}{m_{\Lambda_b}^2}|B|^2 \right ]\label{non-lep-br}, \eea
where $p_{cm}$ is magnitude of the center-of-mass momentum of the
outgoing particles.

For numerical analysis we use the following input parameters. The
masses of the particles, the decay constant of pion and the lifetime
of $\Lambda_b$ baryon are taken from \cite{pdg}. The values of the
effective Wilson coefficients are taken from \cite{rm2001}. The
values of the CKM elements used are $|V_{ub}|=( 3.93  \pm
0.36)\times 10^{-3}$, $|V_{us}|=( 0.2255  \pm 0.0019)$,
$|V_{tb}|=0.999 $, $|V_{ts}|=( 38.7  \pm 2.3)\times 10^{-3}$
\cite{pdg}
 and the weak phase
$\gamma=\left (70_{-21}^{+14} \right )^\circ$
\cite{ckmfitter}.

To evaluate the branching ratio for $\Lambda_b \to \Lambda \pi $
decay  we need to specify the form factors describing $ \Lambda_b
\to \Lambda $ transition. In this analysis we use the values of the
factors from \cite{cdlu} which are evaluated using the light-cone
sum rules. In this approach, the dependence of form factors on the
momentum transfer can be parameterized as \be \xi_i (q^2)=
\frac{\xi_i(0)}{1-a_1 (q^2/m_{\Lambda_b}^2) +a_2
(q^4/m_{\Lambda_b}^4)}\;,\label{form} \ee where $\xi$ denotes the
form factor $g_1$ and $g_2$. The values of the parameters
$\xi_i(0)$, $a_1$ and $a_2$ have been presented in Table-1. The
other form factors can be related to these two as \bea g_1=G_1,~~~~~
g_2=G_2=g_3=G_3. \eea

\begin{table}[t]
\begin{center}
\caption{ Numerical values of the form factors  $g_1$ and $g_2$ and
the parameters $a_1$ and $a_2$ involved in the double fit
(\ref{form}).}
\begin{tabular}{|c|cc|cc|}
\hline \hline
parameter && twist-3 && up to twist-6\\
\hline
$g_1(0)$ && $0.14_{-0.01}^{+0.02} $ && $0.15_{-0.02}^{+0.02} $\\
$a_1 $&& $2.91_{-0.07}^{+0.10} $ && $2.94_{-0.06}^{+0.11} $\\
$a_2$ && $2.26_{-0.08}^{+0.13} $ && $2.31_{-0.10}^{+0.14} $\\
 \hline
$ g_2(0)( 10^{-2} ~{\rm GeV^{-1}}) $ && $-0.47_{-0.06}^{+0.06} $ && $1.3_{-0.4}^{+0.2} $\\
$a_1 $ && $3.40_{-0.05}^{+0.06} $ && $ 2.91_{-0.09}^{+0.12} $\\
$a_2$ && $2.98_{-0.08}^{+0.09} $ && $ 2.24_{-0.13}^{+0.17} $\\
 \hline
 \hline
\end{tabular}
\end{center}
\end{table}
Thus, we obtain the branching ratio for $\Lambda_b \to \Lambda \pi$
mode in the SM as \bea
{\rm Br}(\Lambda_b \to \Lambda \pi ) & = & (6.4 \pm 2.0) \times 10^{-8}~~~~(\rm twist-3)\nn\\
{\rm Br}(\Lambda_b \to \Lambda \pi ) &=& (7.4 \pm 2.3) \times
10^{-8}~~~~(\rm up ~to~ twist-6), \eea where we have assumed $50 \%$
uncertainties due to non-factorizable contributions. It should be
noted that these values  are beyond the reach of the currently
running experiments and hence, observation of this mode will be a
clear signal of new physics.

In the presence of a fourth generation of quarks, there will be
additional contribution due to the $t'$ quark in the electroweak
penguin loops. Furthermore, it should be noted that due to the
presence of $t'$ quark the unitarity condition becomes $\lambda_u +
\lambda_c+\lambda_t+\lambda_{t'}=0$, where $\lambda_q = V_{qb}
V_{qs}^*$.

Thus, in the presence of the fourth generation of quarks the
amplitude for $\Lambda_b \to \Lambda \pi$ will become \bea {\cal
A}(\Lambda_b \to \Lambda \pi^0)  = i  \left (\lambda_u a_2 -
\frac{3}{2}\lambda_t(a_9 - a_7) -\frac{3}{2}\lambda_{t'}(a'_9 -
a'_7) \right )\times \bar u_{\Lambda}(p')( X + Y \gamma_5
)u_{\Lambda_b}(p), \eea where  $X$ and $Y$ are given as \bea
X &=& \frac{G_F}{2}f_\pi \Big(g_1(q^2) (m_{\Lambda_b}-m_\Lambda ) + g_3(q^2) m_\pi^2 \Big),\nn\\
Y &= &\frac{G_F}{2}f_\pi
 \Big(G_1(q^2)(m_{\Lambda_b} + m_{\Lambda}) -G_3(q^2) m_\pi^2 \Big).
\eea
The above amplitude can be represented in a more general way
\bea
{\cal A}(\Lambda_b(p') &\to &  \Lambda(p) \pi^0(q)) =
 i [\bar u_{\Lambda}  (X + Y \gamma_5) u_{\Lambda_b}]\nn\\
&\times &  \lambda_u a_2 \Big(1 +r a \exp(i(\delta +\gamma) - b r'
\exp(i(\delta'+\phi_s+\gamma))\Big), \eea where the parameters $a$,
$b$, $r$, $r'$ and the strong phases $\delta$ and $\delta'$ are
defined as \bea
a&=&|\lambda_t/\lambda_u|,~~~~b=|\lambda_t'/\lambda_u|, ~~~~r=
\frac{3}{2}\left |\frac{a_9-a_7}{a_2}\right |,~~~~~~~
r'=\frac{3}{2}\left |\frac{a_9'-a_7'}{a_2}\right |\nn\\
\delta &= &\arg \left (\frac{a_9-a_7}{a_2}\right ),~~~
\delta'=\arg\left ( \frac{a_9'-a_7'}{a_2} \right ). \eea The weak
phases of the CKM elements are used as : $(-\gamma)$ the phase of
$V_{ub}$,  $\pi$ is the phase of  $V_{ts}$ and $\phi_s$ is the phase
of $\lambda_{t'}$. The decay width for this process can be given by
\bea \Gamma &=&  \frac{p_{cm}}{8 \pi}|\lambda_u a_2|^2 \left [
\frac{(m_{\Lambda_b}+m_\Lambda)^2-m_{\pi}^2}{m_{\Lambda_b}^2}|X|^2
+\frac{(m_{\Lambda_b}-m_{\Lambda})^2 - m_\pi^2}{m_{\Lambda}^2}|Y|^2
\right ] \nn\\
&\times &  \Big[1+a^2 r^2+b^2 r'^2 + 2 a r \cos (\delta +\gamma) -2
b r' \cos (\phi_s +\gamma +\delta') - 2 a b r r' \cos (\phi_s +
\delta' -\delta ) \Big].\nn\\ \eea

For numerical evaluation of the branching ratio we need to know the
values of the new parameters of this model. We use the allowed range
for the new CKM elements as $|\lambda_{t'}|=(0.08 \rightarrow 1.4)
\times 10^{-2}$ and $\phi_s=(0 \rightarrow 80)^\circ$ for
$m_{t'}=400 $ GeV, extracted using the available  observables which
are mediated through $b \to s$ transitions \cite{rm1}. To find out
the values of the QCD parameters $a_9'$ and $a_7'$ we need to
evaluate the new Wilson coefficients $C_{7-10}'$ due to the virtual
$t'$ quark exchange in the loop. The values of these coefficients at
$M_W$ scale can be obtained from the corresponding contribution due
to $t$-quark exchange by replacing the mass of $t$ quark in the
Inami-Lim functions \cite{inami} by $m_{t'}$. These values can then
be evolved to the $m_b$ scale using the renormalization group
equation as discussed in \cite{buras1}. The values of these
coefficients for a representative $t'$ mass mass $m_{t'}=400$ GeV
listed in Table-2.

\begin{table}[t]
\begin{center}
\caption{ Numerical values of the Wilson coefficients $C_i'$ for
$m_{t'}=400$ GeV.}
\begin{tabular}{|c|c|c|c|}
\hline
$C_3'$ & $C_4'$ & $C_5'$ & $C_6'$\\
\hline
$2.06 \times 10^{-2}$&$-3.85 \times 10^{-2}$& $1.02 \times 10^{-2}$ & $-4.43 \times 10^{-2}$\\
\hline  $C_7'$ & $C_8'$ & $C_9'$ & $C_{10}' $\\
\hline   ~$4.453 \times 10^{-3} $~ &~$2.115 \times 10^{-3} $~ &~
$-0.029  $ ~&~$0.006 $ ~\\
 \hline

\end{tabular}
\end{center}
\end{table}

With these inputs the variation of the branching ratio for the
$\Lambda_b \to \Lambda \pi$ with $|\lambda_t'|$ is shown in
Figure-1. From the figure it can be seen that the  branching ratio
is significantly enhanced from its corresponding SM value and it
could be easily accessible in the currently running LHCb experiment.

%%%%%%%%%%%%%%%%%%%%%%%%%%%%%%%%%%%%%%%%%%%%%%%%%%%%%%%%%%%%%%%%%%%%%%%%%%%%%%%

 \begin{figure}[htb]
  \centerline{\epsfysize 3.0 truein \epsfbox{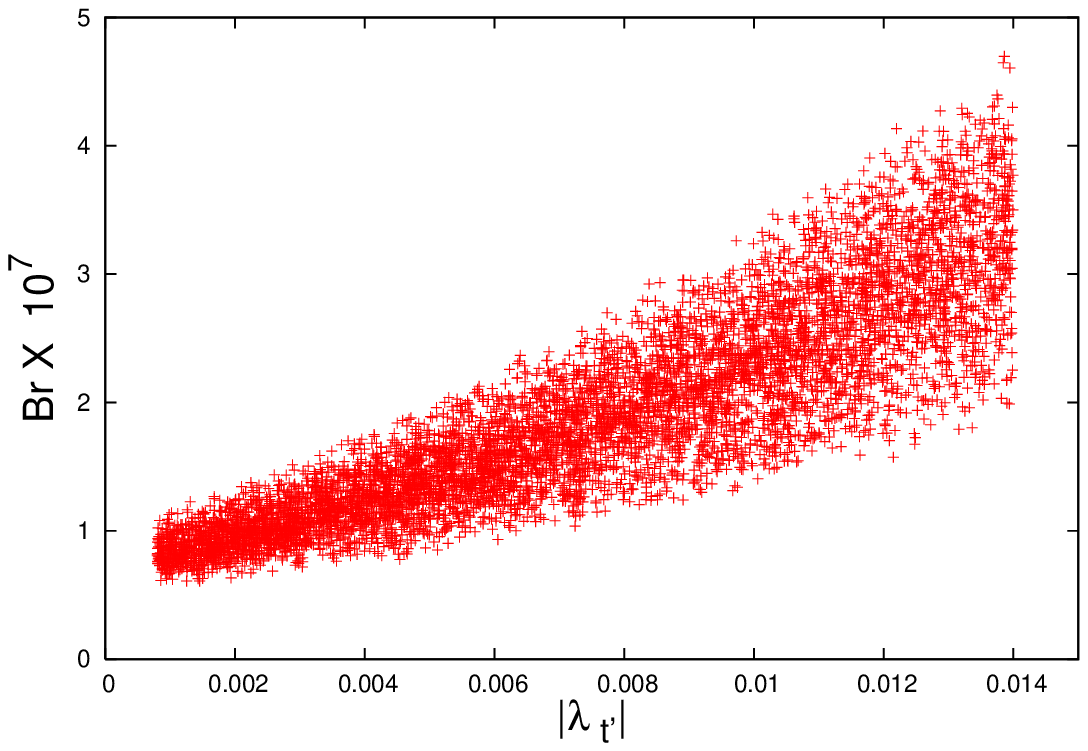}}
\caption{
  The branching ratio   versus $|\lambda_t'| $ for the process  $\lb \to  \ll \pi $. }
  \end{figure}

%%%%%%%%%%%%%%%%%%%%%%%%%%%%%%%%%%%%%%%%%%%%%%%%%%%%%%%%%%%%%%%%%%%%%%%%%%%%%%%%%%%%%%%%%%%%%

Now we will discuss the decay mode $\Lambda_b$ decay mode $\Lambda_b
\to p K^- $, mediated through $b \to s $ transition. In the SM, it
receives contributions from color allowed tree, QCD as well as
electroweak penguins. Its amplitude in the SM is given as
\cite{rm2001} \bea {\cal A}(\Lambda_b \to p K^-) &= & i
\frac{G_F}{\sqrt 2} f_K \bar u_p(p')
\Big[\Big(\lambda_u a_1 -\lambda_t (a_4+a_{10}+(a_6+a_8)R_1\Big)\nn\\
&\times & (g_1(m_K^2) (m_{\Lambda_b}-m_\Lambda)+g_3(m_K^2) m_K^2)\nn\\
&+&\Big(\lambda_u a_1 -\lambda_t (a_4+a_{10}-(a_6+a_8)R_2\Big)\nn\\
&\times & \Big(G_1(m_K^2) (m_{\Lambda_b}+m_\Lambda)-G_3(m_K^2)
m_K^2\Big)\gamma_5\Big] u_{\Lambda_b}(p),\label{pk} \eea where \be
 R_1=\frac{2 m_K^2}{(m_b - m_u)(m_s+m_u)},~~~~~~~R_2=\frac{2 m_K^2}{(m_b+m_u)(m_s+m_u)}.
\ee

From the above amplitude one can obtain the branching ratio using
Eq. (\ref{non-lep-br}). Using the input parameters as discussed
earlier in this section and assuming $50\%$ uncertainties due to
nonfactorizable contributions, we obtain the branching ratio in the
SM \be {\rm Br}(\Lambda_b \to p K^-) = 3.5 \times 10^{-6} \ee which
is lower than the present experimental value ${\rm Br}(\Lambda_b \to
p K^-) = (5.6 \pm 0.8 \pm 1.5) \times 10^{-6}$ \cite{alto}. Here we
have used the form factors for $\Lambda_b \to p$ transitions from
\cite{ztwei}, which are evaluated in the light-front quark model.
The $q^2$ dependence of the form factors are given by the following
three parameters fit as \be
\xi_i(q^2)=\frac{\xi_i(0)}{(1-q^2/m_{\Lambda_b}^2)( 1-a_1
(q^2/m_{\Lambda_b}^2) +a_2 (q^4/m_{\Lambda_b}^4)}\;,\label{form-p}
\ee where the values of the different fit parameters are listed in
Table-3.

\begin{table}[t]
\begin{center}
\caption{ Numerical values of the form factors  $g_1$ and $g_2$ and
the parameters $a_1$ and $a_2$ for $\Lambda_b \to p $ transition
(\ref{form-p}).}
\begin{tabular}{|c|cc|cc|cc|}
\hline
~~~$\xi$~~~ && ~~~$\xi(0) $~~~ && ~~~$a$~~~ &&~~~$ b$~~~\\
\hline
~~~$g_1$~~~ && ~~~$0.1131 $~~~ && ~~~1.70~~~ && ~~~1.60~~~ \\
$g_3$ && $0.0356 $ && 2.5 && 2.57 \\
~~~$G_1$~~~ && ~~~$0.1112 $~~~ && ~~~1.65~~~ && ~~~1.60~~~ \\
$G_3$ && $0.0097 $ && 2.8 && 2.7 \\
 \hline
\end{tabular}
\end{center}
\end{table}

As discussed earlier in the presence of a fourth generation of
quarks the amplitude (\ref{pk}) will receive additional
contributions due to the heavy $t'$ quark in the loop. The modified
amplitude becomes
 \bea
{\cal A}(\Lambda_b \to p K^-) &= & i \frac{G_F}{\sqrt 2} f_K \bar u_p\Big[\Big(
\lambda_u a_1 -\lambda_t (a_4+a_{10}+(a_6+a_8)R_1)\nn\\
&-& \lambda_{t'} (a_4'+a_{10}'+(a_6'+a_8')R_1)\Big)(g_1(m_K^2) (m_{\Lambda_b}-m_\Lambda)+g_3(m_K^2) m_K^2)\nn\\
&+&\Big(\lambda_u a_1 -\lambda_t (a_4+a_{10}-(a_6+a_8)R_2-\lambda_{t'} (a_4'+a_{10}'-(a_6'+a_8')R_2)\Big)\nn\\
&\times & \Big(G_1(m_K^2) (m_{\Lambda_b}+m_\Lambda)-G_3(m_K^2)
m_K^2\Big )\gamma_5\Big] u_{\Lambda_b}. \eea

Now using the values of the new Wilson coefficients $C_{3-10}'$ from
Table-2 and varying the new CKM elements between  $0.0008 \leq
|\lambda_{t'}|\leq 0.014$ and $(0 \leq \phi_s \leq 80)^\circ$, we
present in Figure-2 the variation of Br($\Lambda_b \to p K^-)$ with
$|\lambda_{t'}|$. From the figure it can be seen that the measured
branching ratio can be easily accommodated in this model.
%%%%%%%%%%%%%%%%%%%%%%%%%%%%%%%%%%%%%%%%%%%%%%%%%%%%%%%%%%%%%%%%%%%%%%%%%%%%%%%

 \begin{figure}[htb]
  \centerline{\epsfysize 3.0 truein \epsfbox{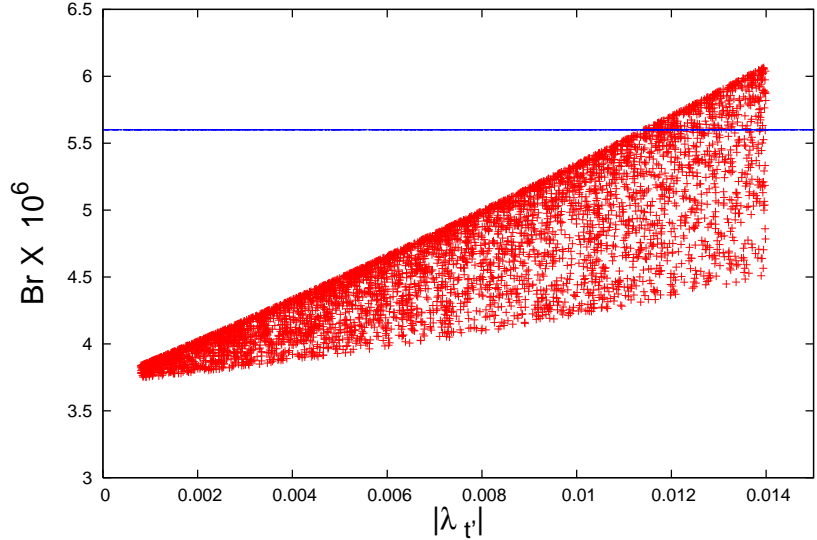}}
\caption{
  The branching ratio   versus $|\lambda_t'| $ for the process  $\lb \to p K^- $,
  where the horizontal line represents the  experimental central value.}
  \end{figure}

%%%%%%%%%%%%%%%%%%%%%%%%%%%%%%%%%%%%%%%%%%%%%%%%%%%%%%%%%%%%%%%%%%%%%%%%%%%%%%%%%%%%%%%%%%%%%
\section{$\Lambda_b \to \Lambda \gamma $ decay width}

In this section we will consider the rare radiative decay $\Lambda_b
\to \Lambda \gamma$ which is induced by the quark level transition $
b \to s \gamma $. The effective Hamiltonian describing $\Lambda_b
\to \Lambda \gamma $ is given as \be {\cal H}_{eff}= -
\frac{4G_F}{\sqrt 2} \lambda_t C_7(m_b) O_7, \ee where $C_7$ is the
Wilson coefficient and $O_7$ is the electromagnetic dipole operator
given as \be O_7= \frac{e}{32 \pi^2}F_{\mu \nu}[ m_b \bar s
\sigma^{\mu \nu} (1+\gamma_5) b + m_s \bar s \sigma^{\mu \nu}
(1-\gamma_5) b ] \ee The expression for calculating the Wilson
coefficient $C_7(\mu)$ is given in \cite{buras2}. The matrix
elements of the various hadronic currents between initial $\lb$ and
the final $\ll$ baryon, which are parameterized in terms of various
form factors as  \bea \langle \ll |\bar s i \sigma_{\mu \nu} q^\nu b
| \lb \rangle & =& \bar u_\ll \Big[ f_1 \gamma_\mu + i
f_2\sigma_{\mu \nu} q^\nu +f_3 q_\mu \Big]
u_{\lb}\;,\nn\\
\langle \ll |\bar s i \sigma_{\mu \nu}\gamma_5 q^\nu b | \lb \rangle
& =& \bar u_\ll \Big[ F_1 \gamma_\mu \gamma_5 + i F_2 \sigma_{\mu
\nu}\gamma_5 q^\nu + F_3 \gamma_5 q_\mu \Big]
u_{\lb}\;,\label{matrix1} \eea These form factors are related to the
previously defined $g_1$ and $g_2$ through \cite{cdlu}
 \bea
&&F_1(q^2)=f_1(q^2)=q^2 g_2(q^2)=q^2 G_2(q^2), \nn\\
&& F_2(q^2) = f_2(q^2) =g_1(q^2)=G_1(q^2). \eea

Thus, one can obtain the decay width of $\Lambda_b \to \Lambda
\gamma $ in the SM as \be \Gamma( \Lambda_b \to \Lambda \gamma) =
\frac{\alpha G_F^2}{32 m_{\Lambda_b}^3 \pi^4} |V_{tb} V_{ts}^*|^2
|C_7|^2 (1-x^2)^3 (m_b^2 +m_s^2) [f_2(0)]^2,\label{rad}
 \ee
where $x=m_{\Lambda}/m_{\Lambda_b}$. Using the input parameters
 as discussed in section II we obtain the
branching ratio in the SM as \be {\rm Br}(\Lambda_b \to \Lambda
\gamma)= (7.93 \pm 2.31) \times 10^{-6}, \ee which is well below the
present experimental upper limit ${\rm Br}(\Lambda_b \to \Lambda
\gamma)< 1.3 \times 10^{-3}$ \cite{pdg}. Now we would like to see
the effect of fourth quark generation on the branching ratio of
$\Lambda_b \to \Lambda \gamma $. In the presence of fourth quark
generation of quarks, the Wilson coefficient $C_7$ will be modified
due to the $t'$ contribution in the loop. Thus the modified
parameter can be given as
 \be
 C_7^{\rm tot}(\mu)=C_7(\mu)+
  \frac{V_{t' b}V_{t's}^*}{V_{tb}V_{ts}^*} C_7'(\mu).
 \ee
where $C_7'$ can be obtained from the expression of $C_7$ by replacing the mass of
$t$ quark by $m_{t'}$. The value of $C_7'$ for $m_{t'}=400$ GeV is found to be
$C_7'=-0.375$.

Thus, in SM4 the branching ratio can be given by Eq. (\ref{rad}) by
replacing $C_7$ by $C_7^{tot}$. Now varying $\lambda_{t'} $ between
$0.0008 \leq |\lambda_t'| \leq 0.0014$ and $\phi_s$ between
$(0-80)^\circ$ we show in Figure -3 the corresponding branching
ratio, where we have included $30\%$ uncertainties due to hadronic
form factors. From the figure it can be seen that the branching
ratio in SM4 has been significantly enhanced from its SM value and
it could be easily accessible it the currently running experiments.

%%%%%%%%%%%%%%%%%%%%%%%%%%%%%%%%%%%%%%%%%%%%%%%%%%%%%%%%%%%%%%%%%%%%%%%%%%%%%%%

 \begin{figure}[htb]
  \centerline{\epsfysize 3.0 truein \epsfbox{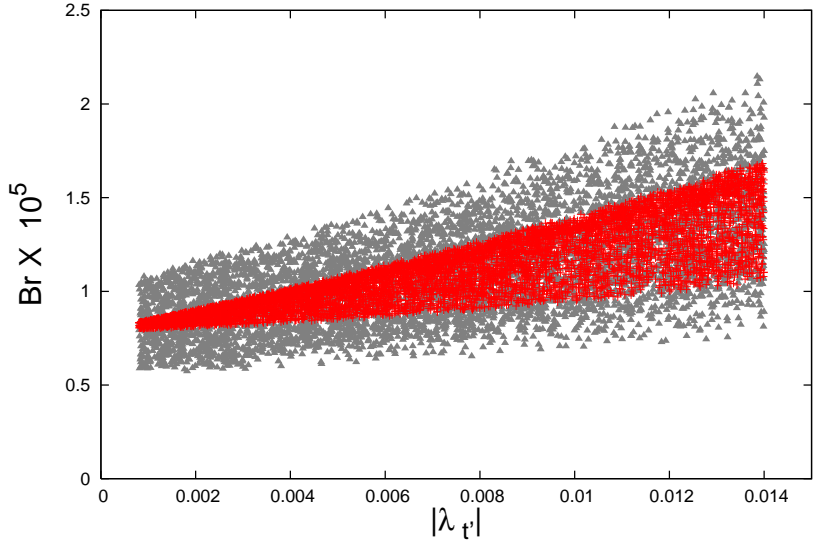}}
\caption{
  The branching ratio  versus $|\lambda_t'| $ for the process  $\lb \to  \ll \gamma $. The
grey bands are due to the $30 \%$ uncertainties in the hadronic form
factors}
  \end{figure}

%%%%%%%%%%%%%%%%%%%%%%%%%%%%%%%%%%%%%%%%%%%%%%%%%%%%%%%%%%%%%%%%%%%%%%%%%%%%%%%%%%%%%%%%%%%%%

\section{$\Lambda_b \to \Lambda l^+ l^- $ decays}

The decay process $\lb \to \ll~ l^+ l^-$ is described by the quark
level transition $ b \to s l^+ l^-$. These processes are extensively
studied in the literature \cite{aliev} in various beyond the
standard model scenarios. The effective Hamiltonian describing these
processes can be given as \cite{buras1} \bea {\cal H}_{eff} &=
&\frac{ G_F~ \alpha}{\sqrt 2 \pi}~ V_{tb} V_{ts}^*~\Big[
C_9^{eff}(\bar s \gamma_\mu L b)(\bar l \gamma^\mu l) \nn\\
&+& C_{10}(\bar s \gamma_\mu L b)(\bar l \gamma^\mu \gamma_5 l) -2 C_7^{eff}
 m_b(\bar s i \sigma_{\mu \nu} \frac{q^\mu}{q^2} R b)
(\bar l \gamma^\mu l) \Big]\;,\label{ham}
\eea
where $q$ is the momentum transferred to the lepton pair, given as
$q=p_-+p_+$, with $p_-$ and $p_+$ are the momenta of the leptons $l^-$
and $l^+$ respectively. $L,R=(1 \pm \gamma_5)/2$ and $C_i$'s are the Wilson
coefficients evaluated at the $b$ quark mass scale. The values of these
coefficients in NLL order are
$C_7^{eff}=-0.31\;,~~C_9=4.154\;,~~C_{10}=-4.261$ \cite{beneke}.

The coefficient $C_9^{eff}$ has a perturbative part and a
resonance part which comes
from the long distance effects due to the conversion of the real
 $c \bar c$ into the lepton pair $l^+ l^-$. Therefore, one can write it as
\be
C_9^{eff}=C_9+Y(s)+C_9^{res}\;,
\ee
where $s=q^2$ and the function $Y(s)$ denotes the perturbative part coming
from one loop matrix elements  of the four quark operators and
is given by \cite{buras1}
\bea
Y(s)&=& g(m_c,s)(3 C_1+C_2+3C_3+C_4+3C_5+C_6) -\frac{1}{2} g(0,s)
(C_3+3C_4)\nn\\
&-&\frac{1}{2} g(m_b,s)(4 C_3+4 C_4+3 C_5 +C_6)
+ \frac{2}{9}(3 C_3+C_4+3C_5+C_6)\;,
\eea
where
\bea
g(m_i,s) &=& -\frac{8}{9} \ln(m_i/m_b^{pole}) + \frac{8}{27}+\frac{
4}{9}y_i -\frac{2}{9}(2+y_i)\sqrt{|1-y_i|}\nn\\
&\times & \biggr\{\Theta(1-y_i)\biggr[\ln\left (
\frac{1+\sqrt{1-y_i}}{1-\sqrt{1-y_i}}\right )-i \pi \biggr]
+\Theta(y_i-1)2 \arctan \frac{1}{\sqrt{y_i-1}} \biggr\}\;, \eea with
$y_i=4 m_i^2/s$. The values of the coefficients $C_i$'s in NLL order
are taken from \cite{beneke}.

The long distance resonance effect is given as \cite{res} \bea
C_9^{res}= \frac{3 \pi}{\alpha^2}(3 C_1+C_2+3C_3+C_4+3C_5+C_6)\sum_{
V_i=\psi(1S), \cdots, \psi(6S) } \kappa_{V_i}\frac{m_{V_i}
\Gamma(V_i \to l^+ l^-)}{m_{V_i}^2 -s -i m_{V_i}\Gamma_{V_i}}\;.
\eea The phenomenological parameter $\kappa$ is taken to be 2.3, so
as to reproduce the correct branching ratio of ${\rm Br}(B \to
J/\psi K^* l^+ l^-)={\rm Br}(B \to J/\psi K^*) {\rm Br}(J/\psi \to
l^+ l^-)$.

The matrix elements of the various hadronic currents in (\ref{ham})
between initial $\lb$ and the final $\ll$ baryon, which are
parameterized in terms of various form factors as defined in Eqs.
(\ref{matrix}) and (\ref{matrix1}). Thus, using these matrix
elements, the transition amplitude can be written as \bea {\cal
M}(\lb \to \ll l^+ l^-) &=& \frac{G_F~ \alpha}{\sqrt 2 \pi} V_{tb}
V_{ts}^* \Biggr[ \bar l \gamma_\mu l \Big\{ \bar u_\ll
\Big(\gamma^\mu (A_1 P_R+B_1 P_L)+ i \sigma^{\mu \nu} q_\nu
(A_2 P_R +B_2 P_L) \Big)u_{\lb} \Big\}\nn\\
&+&\bar l \gamma_\mu \gamma_5 l \Big\{
\bar u_\ll \Big(\gamma^\mu  (D_1 P_R+E_1 P_L)+ i \sigma^{\mu \nu} q_\nu
(D_2 P_R +E_2 P_L)\nn\\
&+& q^\mu(D_3 P_R +E_3 P_L) \Big)u_{\lb} \Big\}\Biggr]\;,\label{e1}
\eea
where the various parameters $A_i,~B_i$ and $D_j,~E_j$
($i=1,2$ and $j=1,2,3$) are defined as
\bea
A_i &=& \frac{1}{2}C_9^{eff}(g_i-G_i)-\frac{C_7 m_b}{q^2}
(f_i+F_i)\;,\nn\\
B_i &=& \frac{1}{2}C_9^{eff}(g_i+G_i)-\frac{C_7 m_b}{q^2}
(f_i-F_i)\;,\nn\\
D_j &=& \frac{1}{2}C_{10} (g_j-G_j)\;,~~~~ E_j = \frac{1}{2}C_{10}
(g_j+G_j)\;. \eea We will consider here  the case when the final
$\ll$ baryon is unpolarized. The physical observables in this case
are the differential decay rate and the forward backward rate
asymmetries. From the transition amplitude (\ref{e1}), one can
obtain double differential decay rate  \cite{rm2006} as \bea
\frac{d^2 \Gamma}{d\hat s~ dz}=\frac{G_F^2~ \alpha^2}{2^{12} \pi^5}~
|V_{tb} V_{ts}^*|^2~m_{\lb}~ v_l~ \lambda^{1/2}(1, r, \hat s)~ {\cal
K}(s ,z)\;, \label{e2} \eea where $\hat s=s/m_{\lb}^2$, $z=\cos
\theta $, the angle between $p_{\lb}$ and  $p_{+}$  in the center of
mass frame of $l^+ l^-$ pair, $v_l=\sqrt{1-4 m_l^2/ s}$ and
$\lambda(a,b,c)=\sqrt{a^2+b^2+c^2-2(ab+bc+ca)}$ is the usual
triangle function. The function ${\cal K}(s, z)$ is given as \be
{\cal K}(s,z)={\cal K}_0(s)+z~ {\cal K}_1(s)+z^2~{\cal K}_2(s)\;,
\ee with \bea
{\cal K}_0(s) &=& 32 m_l^2 m_{\lb}^2\s1(1+r -\s1)(|D_3|^2+|E_3|^2)\nn\\
&+&
64 m_l^2 m_{\lb}^3(1-r -\s1)Re(D_1^*E_3+D_3 E_1^*)
+64  m_{\lb}^2 \sqrt{r} (6 m_l^2-\s1 m_{\lb}^2)Re(D_1^*E_1)
\nn\\
&+&64 m_l^2 m_{\lb}^3 \sqrt{r} \Big(
2 m_{\lb} \s1 Re(D_3^*E_3)+(1-r+\s1)Re(D_1^* D_3+ E_1^* E_3)\Big)\nn\\
&+&32 m_{\lb}^2 (2 m_l^2+m_{\lb}^2 \s1)\Big((1-r +\s1)m_{\lb}
\sqrt{r}Re(A_1^*A_2+B_1^* B_2)\nn\\
&-& m_{\lb}(1-r-\s1) Re(A_1^* B_2+A_2^* B_1)-2 \sqrt{r}\Big[Re(A_1^* B_1)
+m_{\lb}^2 \s1 Re(A_2^* B_2)\Big]\Big)\nn\\
&+& 8 m_{\lb}^2\Big(4 m_l^2(1+r-\s1)+m_{\lb}^2[(1-r)^2-\s1^2]\Big)
\Big(|A_1|^2+|B_1|^2\Big)\nn\\
&+& 8 m_{\lb}^4\Big(4 m_l^2[\lambda+(1+r-\s1)\s1]+m_{\lb}^2
\s1[(1-r)^2-\s1^2]\Big)
\Big(|A_2|^2+|B_2|^2\Big)\nn\\
&-& 8 m_{\lb}^2\Big(4 m_l^2(1+r-\s1)-m_{\lb}^2[(1-r)^2-\s1^2]\Big)
\Big(|D_1|^2+|E_1|^2\Big)\nn\\
&+& 8 m_{\lb}^5 \s1 v_l^2 \Big(-8 m_{\lb} \s1 \sqrt{r}
Re(D_2^* E_2)+4 (1-r+\s1)\sqrt{r}Re(D_1^* D_2+E_1^* E_2)\nn\\
&-&4(1-r -\s1) Re(D_1^* E_2+D_2^* E_1)
+m_{\lb}[(1-r)^2-\s1^2]
\Big[|D_2|^2+|E_2|^2\Big]\Big)\;,
\eea
\bea
{\cal K}_1(s) &=& -16  m_{\lb}^4\s1 v_l \sqrt{\lambda}
\Big\{ 2 Re(A_1^* D_1)-2Re(B_1^* E_1)\nn\\
&+& 2m_{\lb}
Re(B_1^* D_2-B_2^* D_1+A_2^* E_1-A_1^*E_2)\Big\}\nn\\
&+&32 m_{\lb}^5 \s1~ v_l \sqrt{\lambda} \Big\{
m_{\lb} (1-r)Re(A_2^* D_2 -B_2^* E_2)\nn\\
&+&
\sqrt{r} Re(A_2^* D_1+A_1^* D_2-B_2^*E_1-B_1^* E_2)\Big\}\;,
\eea
and
\bea
{\cal K}_2(s)&= & 8m_{\lb}^6 v_l^2~ \lambda \s1~ \Big (
(|A_2|^2+|B_2|^2+|D_2|^2+|E_2|^2\Big)\nn\\
&-&8m_{\lb}^4 v_l^2
~\lambda~\Big(|A_1|^2+|B_1|^2+|D_1|^2+|E_1|^2\Big)\;. \eea The
dilepton mass spectrum can be obtained from (\ref{e2}) by
integrating out the angular dependent parameter $z$ which yields \be
\left (\frac{d \Gamma}{d s}\right )_0= \frac{G_F^2~ \alpha^2}
{2^{11} \pi^5 m_{\lb}}~ |V_{tb}V_{ts}^*|^2 v_l~
\sqrt{\lambda}~\Big[{\cal K}_0(s)+ \frac{1}{3} {\cal
K}_2(s)\Big]\;,\label{br1} \ee where $\lambda$ is the short hand
notation for $\lambda(1, r, \hat s)$. The limits for $s$ is \be 4
m_l^2 \leq s\leq (m_{\lb}-m_\ll)^2\;. \ee

Apart from the branching ratio in semileptonic decay, there are also
other observables which are sensitive to new physics contribution in
$b \to s $ transition. One such observable is the forward backward
asymmetry ($A_{FB}$), of leptons which is also a very powerful tool
for looking for new physics. The normalized forward-backward
asymmetry is obtained by integrating the double differential decay
width ($d^2 \Gamma/d \s1 dz)$ with respect to the angular variable
$z$ \be A_{FB}(s)=\frac{\ds{\int_0^1 \frac{d^2 \Gamma}{d \s1
dz}dz-\int_{-1}^0 \frac{d^2 \Gamma}{d \s1 dz}dz}} {\ds{\int_0^1
\frac{d^2 \Gamma}{d \s1 dz}dz+\int_{-1}^0 \frac{d^2 \Gamma}{d \s1
dz}dz}}\;. \ee Thus one obtains from (\ref{e2}) \be
A_{FB}(s)=\frac{{\cal K}_1(s)}{{\cal K}_0(s)+{\cal
K}_2(s)/3}\;.\label{fb} \ee The FB asymmetry becomes zero for a
particular value of dilepton invariant mass. Within the SM, the zero
of $A_{FB}(s)$ appears in the low $q^2$ region, sufficiently away
from the charm resonance region and hence can be predicted
precisely. The position of the zero value of $A_{FB}$ is very
sensitive to the presence of new physics.

For numerical evaluation we use the input parameters as presented in
the previous sections. The quark masses (in GeV) used are $m_b$=4.6,
$m_c$=1.5,  $\alpha=1/128$ and the weak mixing angle $\sin^2
\theta_W=0.23$. The variation of differential branching ratios
(\ref{br1}) and the forward backward asymmetries (\ref{fb}) for the
processes $\Lambda_b \to \Lambda \mu^+ \mu^-$ and $\Lambda_b \to
\Lambda \tau^+ \tau^-$ in the standard model are shown in Figures-4
and 5 respectively.

%%%%%%%%%%%%%%%%%%%%%%%%%%%%%%%%%%%%%%%%%%%%%%%%%%%%%%%%%%%%%%%%%%%%%%%%%%%%
\begin{figure}[htb]
\includegraphics[width=8.0 cm,height=6.0cm,clip]{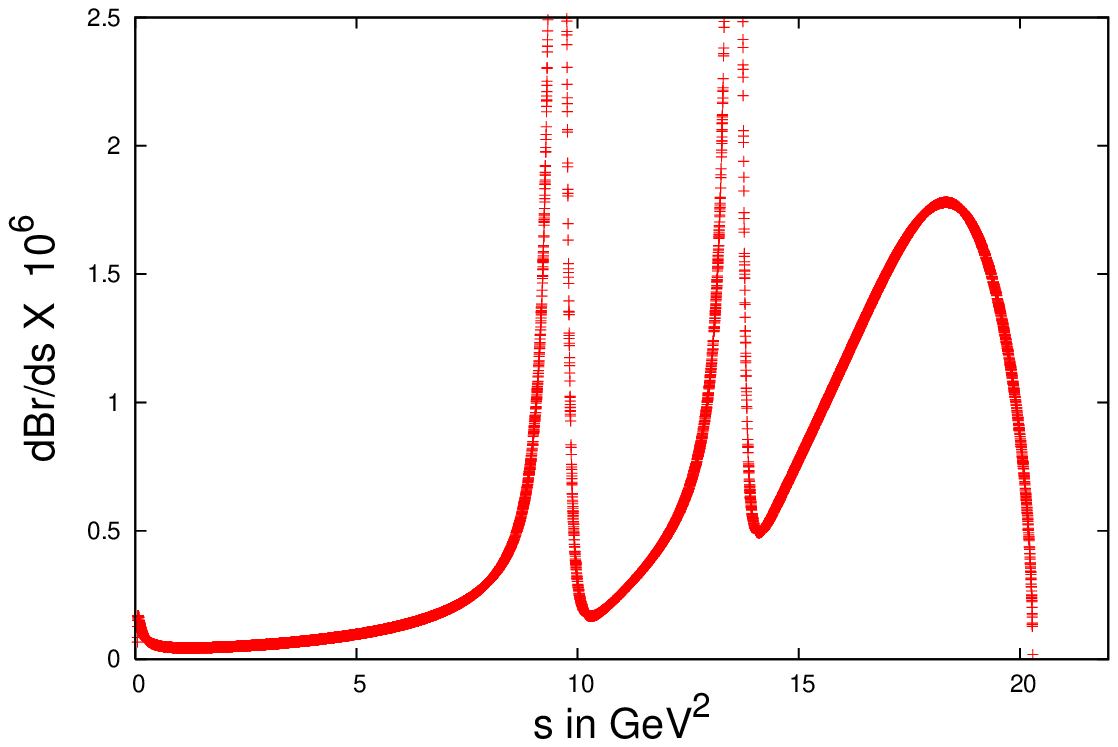}
\hspace{0.2cm}
\includegraphics[width=8.0 cm,height=6.0cm,clip]{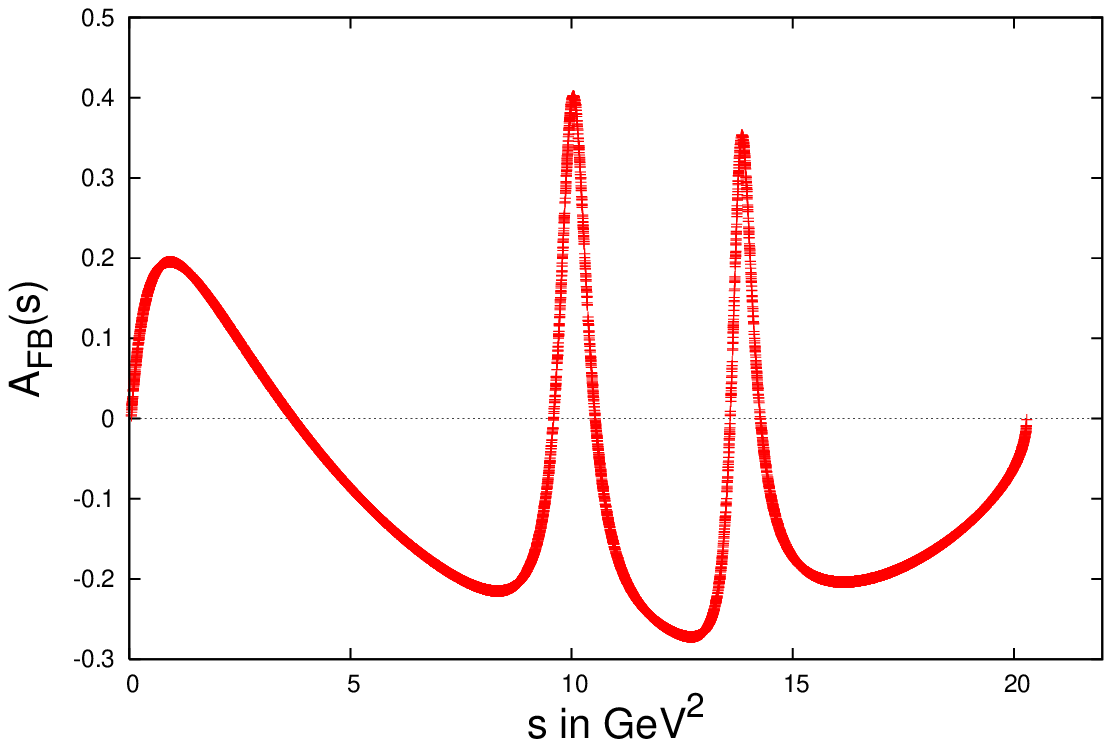}
\caption{The differential branching ratio $d$Br/$ds$ versus $s$
(left panel) and the forward backward asymmetry ($A_{FB}(s)$) versus
$s$ (right panel) for the process $\Lambda_b \to \Lambda \mu^+
\mu^-$.}
\end{figure}
%%%%%%%%%%%%%%%%%%%%%%%%%%%%%%%%%%%%%%%%%%%%%%%%%%%%%%%%%%%%%%%%%%%%%%%%%%%%%%%
\begin{figure}[htb]
\includegraphics[width=8.0 cm,height=6.0cm,clip]{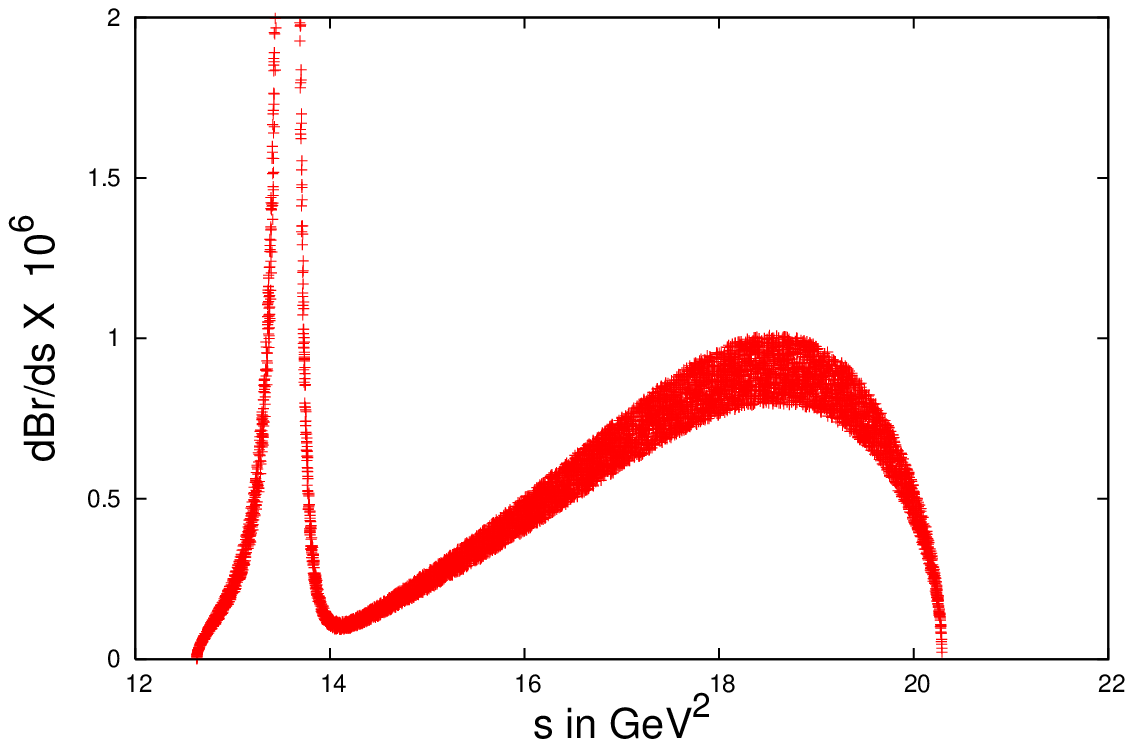}
\hspace{0.2cm}
\includegraphics[width=8.0cm,height=6.0cm,clip]{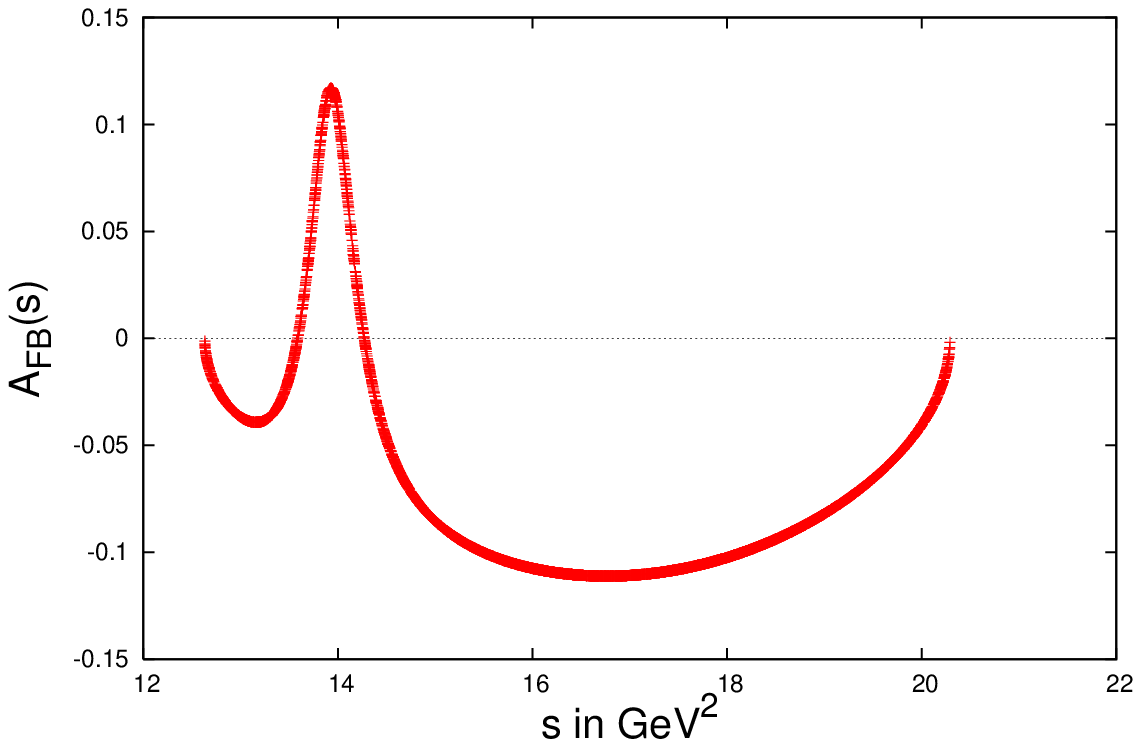}
\caption{Same as Figure-4 for the process $\Lambda_b \to \Lambda
\tau^+ \tau^-$.}
\end{figure}
%%%%%%%%%%%%%%%%%%%%%%%%%%%%%%%%%%%%%%%%%%%%%%%%%%%%%%%%%%%%%%%%%%%%%%%%%%%%%%%

As discussed earlier in the presence of fourth generation, the
Wilson coefficients $C_{7,9,10}$ will be modified due to  the new
contributions arising from the virtual $t'$ quark in the loop. Thus,
these coefficients will be modified as \bea
C_7^{\rm tot}(\mu) &=& C_7(\mu) + \frac{\lambda_{t'}}{\lambda_t} C_7'(\mu),\nn\\
C_9^{\rm tot}(\mu)&= &C_9(\mu) + \frac{\lambda_{t'}}{\lambda_t} C_9'(\mu),\nn\\
C_{10}^{\rm tot}(\mu)&= & C_{10}(\mu) +
\frac{\lambda_{t'}}{\lambda_t} C_{10}'(\mu). \eea The new
coefficients $C_{7,9,10}'$ can be calculated at the $M_W$ scale by
replacing the $t$-quark mass by $m_t'$ in the loop functions.  These
coefficients then to be evolved to the $b$ scale using the the
renormalization group equation as discussed in \cite{buras1}. The
values of the new Wilson coefficients at the $m_b$ scale for
$m_{t'}=400$ GeV is given by $C_7'(m_b)=-0.355$, $C_9'(m_b)=5.831$
and $C_{10}'=-17.358$.

Thus, one can obtain the differential branching ratio and the
forward backward asymmetry in SM4 by replacing $C_{7,9,10}$ in Eqs
(\ref{br1}) and (\ref{fb}) by $C_{7,9,10}^{\rm tot}$. Using the
values of the $|\lambda_t'|$ and $\phi_s$ for $m_{t'}=400$ GeV,
differential branching ratio and the forward  backward asymmetry for
$\Lambda_b \to \Lambda \mu^+ \mu^-$  is presented in Figure-6, where
we have not considered the contributions from intermediate
charmonium resonances. From the figure it can be seen that the
differential branching ratio of  this mode is significantly enhanced
from its corresponding SM value whereas the forward backward
asymmetry is slightly reduced with respect to its SM value. However,
the zero-position of the FB asymmetry remains unchanged the fourth
quark generation model. Similarly for the process $\Lambda_b \to
\Lambda \tau^+ \tau^-$ as seen from Figure-7, the branching ratio
significantly enhanced from its SM value whereas the FB asymmetry
remains almost unaffected in the SM4.

%%%%%%%%%%%%%%%%%%%%%%%%%%%%%%%%%%%%%%%%%%%%%%%%%%%%%%%%%%%%%%%%%%%%%%%%%%%%%%%%
\begin{figure}[htb]
\includegraphics[width=7.5cm,height=5.5cm,clip]{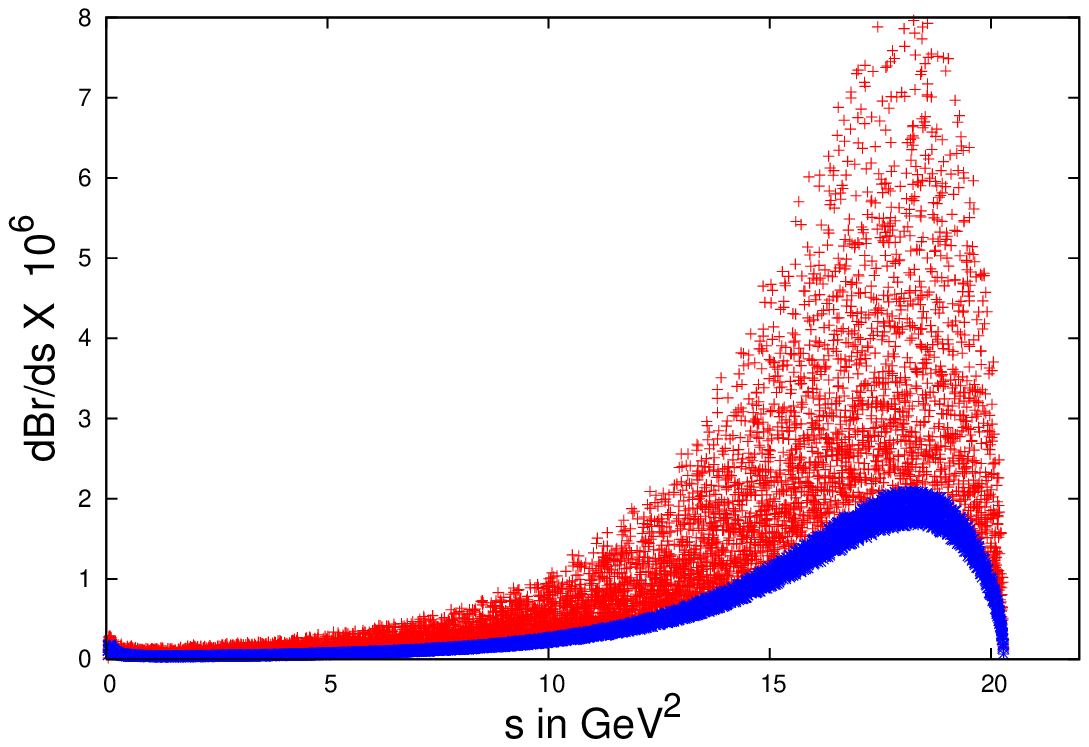}
\hspace{0.2cm}
\includegraphics[width=7.5cm,height=5.5cm,clip]{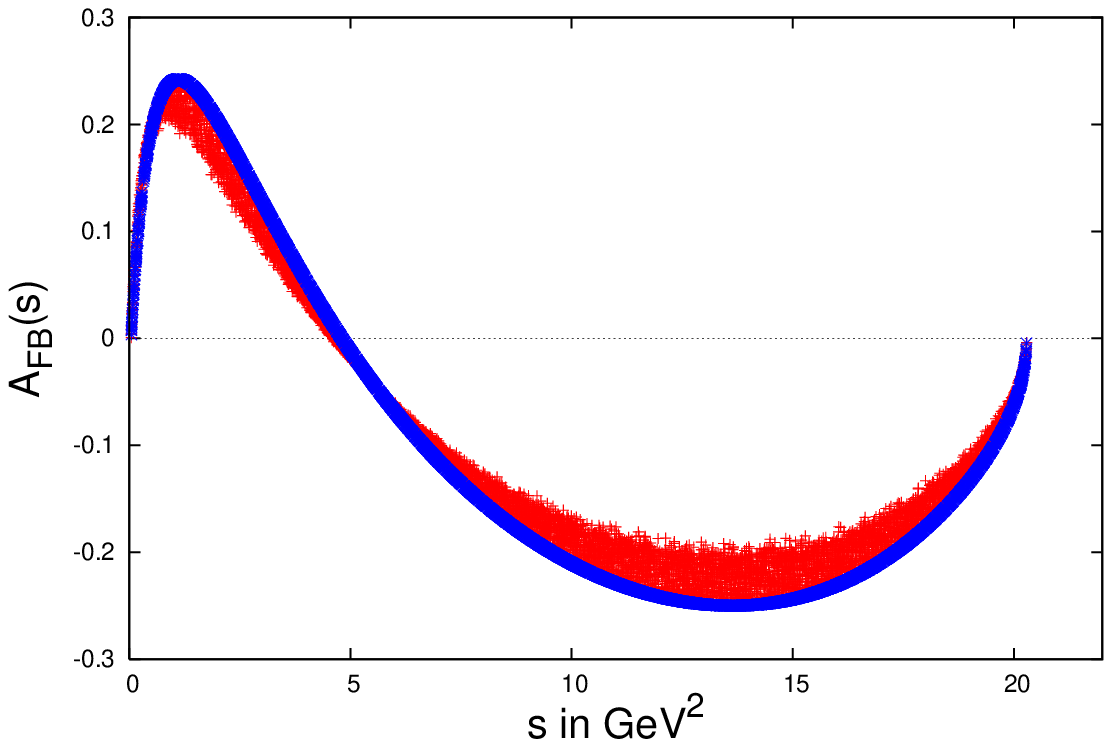}
\caption{Variation of the differential branching ratio (left panel)
and the forward-backward asymmetry (right panel) with respect to the
momentum transfer $s$ for the process $\Lambda_b \to \Lambda \mu^+
\mu^-$, in fourth quark generation model (red regions) whereas the
corresponding SM values are shown by blue regions. }
\end{figure}
%%%%%%%%%%%%%%%%%%%%%%%%%%%%%%%%%%%%%%%%%%%%%%%%%%%%%%%%%%%%%%%%%%%%%%%%%%%%%%%%%%%
\begin{figure}[htb]
\includegraphics[width=8.0cm,height=6.0cm,clip]{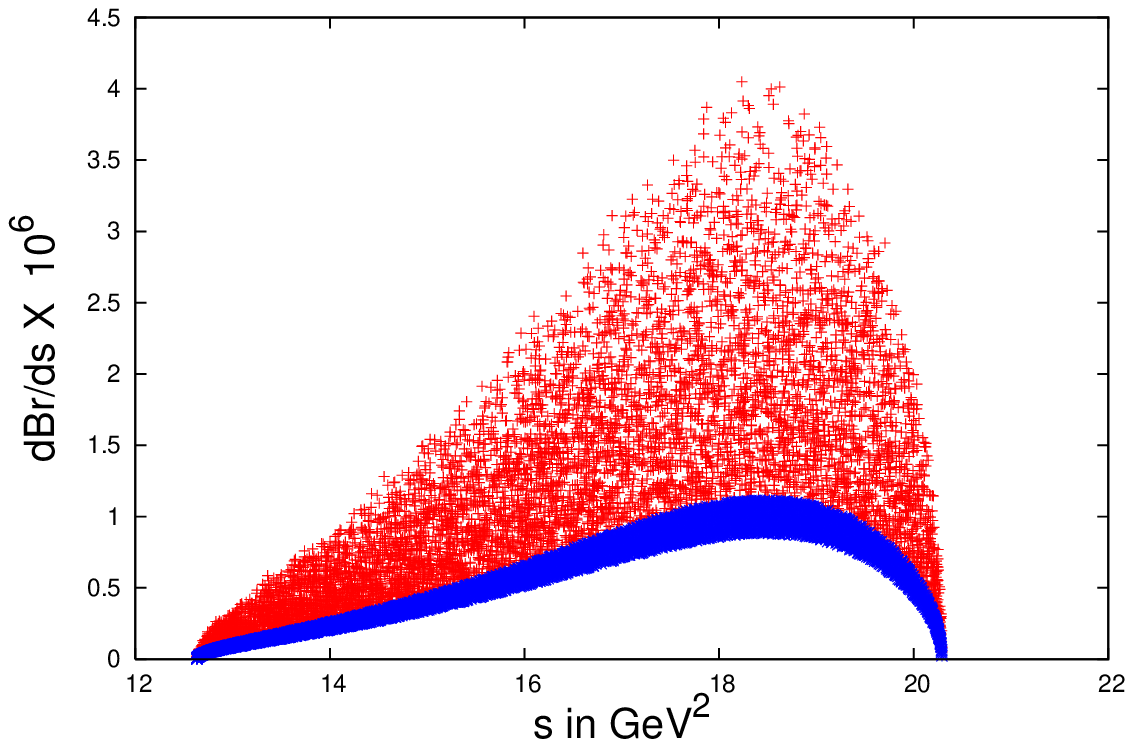}
\hspace{0.2cm}
\includegraphics[width=8.0 cm,height=6.0 cm,clip]{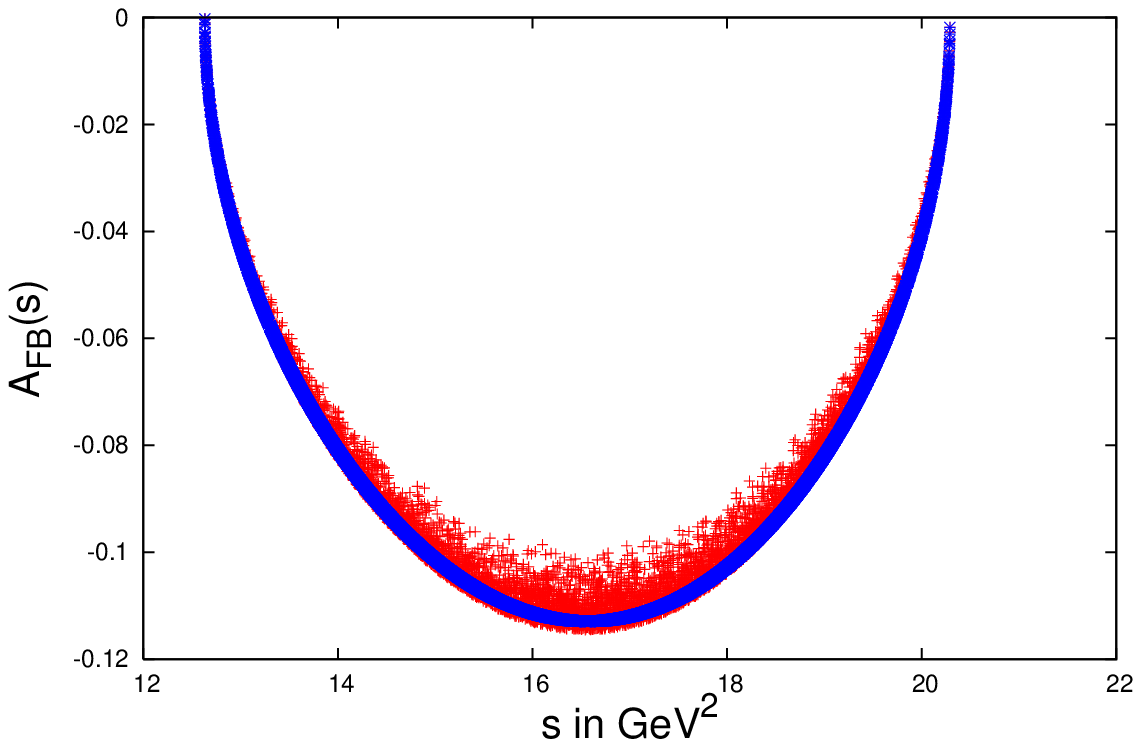}
\caption{Same as Figure-6 for the process $\Lambda_b \to \Lambda
\tau^+ \tau^-$.}
\end{figure}
%%%%%%%%%%%%%%%%%%%%%%%%%%%%%%%%%%%%%%%%%%%%%%%%%%%%%%%%%%%%%%%%%%%%%%%%%%%%%%%%%%%

\begin{table}
\begin{center}
\caption{The branching ratios (in units of $10^{-6}$)
for various decay processes.}
\vspace*{0.3 true cm}
\begin{tabular}{|c|c|c|}
\hline \hline ~Decay modes~ &~  $ {\rm Br}^{\rm SM} $ ~& ~$ {\rm
Br}^{\rm SM4}
$ ~ \\
\hline

$~\lb \to \ll \mu^+ \mu^- $~ & ~13.25 ~ & ~($14.7 \to 53.5$) ~\\
~$\lb \to \ll \tau^+ \tau^- $ ~& ~3.83 ~& ~( $4.3 \to 16.0)~$  \\
\hline
\hline
\end{tabular}
\end{center}
\end{table}

We now proceed to calculate the  total decay rates for $\lb \to \ll~
l^+ l^-$ for which it is necessary to eliminate the backgrounds
coming from the resonance regions. This can be done by by using the
following veto windows so that the backgrounds coming from the
dominant resonances $\lb \to \ll J/\psi (\psi^\prime)$ with
$J/\psi(\psi^\prime)\to l^+ l^-$ can be eliminated,
\begin{eqnarray*}
\lb \to \ll~ \mu^+ \mu^-:&&m_{J/\psi}-0.02<
m_{\mu^+ \mu^-}<m_{J/\psi}+0.02;\nn\\
:&&
 m_{\psi^\prime}-0.02<m_{\mu^+ \mu^-}<m_{\psi^\prime}+0.02 \nn\\
\lb \to  \ll~ \tau^+ \tau^-:&&
m_{\psi^\prime}-0.02<m_{\tau^+ \tau^-}<m_{\psi^\prime}+0.02 \;.
\end{eqnarray*}
Using these veto windows we obtain the branching ratios for
semileptonic rare $\lb$ decays which are presented in Table-4. It is
seen from the table that the branching ratios obtained in the model
in the fourth quark generation model are reasonably enhanced from
the corresponding SM values and could be observed in the LHCb
experiment.

%%%%%%%%%%%%%%%%%%%%%%%%%%%%%%%%%%%%%%%%%%%%%%%%%%%%%%%%%%%%%%%%%%%
\section{Conclusion}

In this paper we have studied several rare  decays of $\Lambda_b$
baryon, i.e., $\Lambda_b \to \Lambda \pi$, $\Lambda_b \to p K^-$,
$\Lambda_b \to \Lambda \gamma$ and $\Lambda_b \to \Lambda l^+ l^-$
in the fourth quark generation model. This model is a very simple
extension of the standard model with three generations and it
provides a simple explanation for the several indications of new
physics that have been observed involving CP asymmetries in the $B$,
$B_s$ decays for $m_t'$ in the range of (400-600) GeV. We found that
in this model the branching ratios of the various decay modes
considered here ($\Lambda_b \to \Lambda \pi$, $\Lambda_b \to p K^-$,
$\Lambda_b \to \Lambda \gamma$ and $\Lambda_b \to \Lambda l^+ l^-$)
are significantly enhanced from their corresponding SM values.
However the forward backward asymmetries in the $\lb \to \ll~ l^+
l^-$ processes do not differ much  from those of the SM
expectations. The zero-point of the $F_{AB}$ for  $\lb \to \ll~ l^+
l^-$ process is also found to be unaffected in this model.

{\bf Acknowledgments}

The work of RM was partly supported by the Department of Science and
Technology, Government of India, through Grant No.
SR/S2/RFPS-03/2006.

%**************************************************************************

\end{document}